%% file: paper.tex
\documentclass{svproc}

\usepackage{url}

\input{packages.tex}

\begin{document}

\mainmatter              
\input{content_R1_arXiv}
\end{document}

%% file: packages.tex
\usepackage{xcolor}
\usepackage{color}
\usepackage{graphicx}
\usepackage{comment}

\usepackage{caption}
\usepackage{amsmath}
\usepackage[extra]{tipa}

\usepackage{multirow}
\usepackage{amsfonts}
\usepackage{makecell}
\usepackage{float} 
\usepackage{enumitem}

\usepackage{tikz}
\usetikzlibrary{arrows.meta,positioning,fit}

\usepackage[table]{xcolor}
\definecolor{tblhead}{gray}{0.93}
\definecolor{tblblock}{gray}{0.97}
\definecolor{tblhi}{gray}{0.90}

\usepackage{booktabs}
\usepackage{makecell}
\usepackage{multirow}
\usepackage{rotating}

\usepackage[normalem]{ulem} 

%% file: content_R1_arXiv.tex

\title{Maximum-Projection-Based Bayesian Optimization Utilizing Sensitivity Analysis for High-Efficiency Radial Turbine Design with Scarce Data}
\titlerunning{Max.-Projection Turbine Design: Sensitivity-Guided Optimization}  
%
\author{Eric Diehl\inst{1} \and Adem Tosun\inst{2}
 \and Dimitrios Loukrezis\inst{3}}
\authorrunning{Diehl, E., Tosun, A., Loukrezis, D.} 
%
\tocauthor{Eric Diehl, Adem Tosun, and Dimitrios Loukrezis}
\institute{Siemens AG, Foundational Technologies, Munich, Germany\\
\email{eric.diehl@siemens.com}\\
\and
Institute of Thermal Turbomachinery and Machinery Laboratory, University of Stuttgart, Stuttgart, Germany
\and
Centrum Wiskunde \& Informatica, Amsterdam, The Netherlands
}

\maketitle              

\begin{abstract}
We propose a data-efficient workflow to optimize the efficiency of a radial turbine design under a strict budget of high-fidelity computational fluid dynamics simulations. 
Assuming anisotropic parameter impact, we use a maximum-projection initial experimental design to ensure space-filling and strong projection properties on low-dimensional subspaces. Bayesian optimization is performed using Gaussian process surrogates with an upper confidence bound acquisition function. 
In parallel, polynomial chaos expansions provide variance-based global sensitivity analysis metrics, which allow to identify a reduced subspace with the most influential parameters, wherein the optimization is continued. 
Turbine efficiency is increased from 85.77\% initially to 91.77\% at the end of the workflow, with a total budget of 330 simulations.

\keywords{Bayesian optimization, maximum projection design, sensitivity analysis, surrogate modeling, radial turbine design}
\end{abstract}


%

%
%
\section{Introduction}
Propulsion systems based on polymer electrolyte membrane fuel cells (PEMFCs) have recently gained attention for their potential integration in short- to medium-range aircraft.  
The application of PEMFCs at elevated heights requires an air-management system, which provides the fuel cell with air at the right pressure, temperature, and humidity for efficient operation. 
A turbine can be used to expand the exhaust gases of the fuel cell and reduce the power consumption of the compressor, thus increasing the overall efficiency of the system \cite{lueck_2022}. 
The interdependence of the different subsystems makes an integrated system design necessary, where fast evaluations of the achievable efficiencies are crucial for the turbo-components \cite{stoewer_2025}. 

Previous studies regarding turbines in PEMFC air-management systems for automotive applications have shown the potential of dropwise condensation, which impacts turbine efficiency and performance \cite{tosun_2024,wittmann_2022,wittmann_2021}. 
Due to the high pressure ratios in aircraft applications, an even more pronounced impact of condensation is expected. Therefore, the influence of condensation on the turbine characteristics needs to be taken into account for reliable component design. 
However, including condensation phenomena in 3D computational fluid dynamics (CFD) simulations drastically increases the computational cost, especially if phase-change is taken into account. 
In turn, design space exploration and optimization workflows become too time-consuming, contrasting the need for fast responses in the iterative PEMFC system design.
State-of-the-art radial turbine design approaches usually employ an initial experimental design (ED) for design space exploration, followed by an adaptive sampling approach for optimization. 
Therein, data-driven surrogate models are often used for design space exploration, reserving high-fidelity CFD model evaluations for design validation \cite{ali_2025,fuhrer_2024,mueller_2013}. 
However, the number of CFD evaluations necessary is typically much higher than justifiable by the available compute resources. 

The aim of this study is to generate an end-to-end optimization workflow based on efficient surrogate modeling techniques, which shall be able to provide an optimized radial turbine design with the minimum number of high-fidelity data samples. 
To that end, we combine space-filling EDs based on maximum projection (MaxPro) \cite{joseph2015maximum}, which are adaptively enriched using Bayesian optimization (BO) \cite{wang2023recent}. 
Simultaneously, we employ global sensitivity analysis (GSA) \cite{saltelli2008global} to reduce the design parameter space and thus mitigate the computational demand. 
Surrogate modeling is a connecting theme, as it is employed in both BO and GSA.
These components are combined in an end-to-end workflow, which is directly applicable to optimization settings dominated by computationally expensive simulations. 
At the same time, it is modular in the sense that individual components can be exchanged depending on the analyst's objective, while preserving the overall workflow structure.

The main challenge is to distribute the very limited CFD simulation budget such that optimization results improve as fast as possible. 
BO builds on probabilistic surrogate models, typically Gaussian processes (GPs), to address the exploration-exploitation trade-off in global optimization \cite{srinivas2009gaussian}, and has become a standard tool for data-efficient optimization in high-cost, black-box simulation settings \cite{wang2023recent,jones1998efficient,shahriari2015taking}.
The quality of the GP model depends crucially on the initial ED, which must feature robust space-filling properties. 
The MaxPro technique is employed here, which additionally provides good \emph{projection} properties, meaning that MaxPro EDs remain sufficiently space-filling if projected to a lower-dimensional subspace \cite{joseph2015maximum}. 
This is an important property for the GSA-based input dimension reduction discussed in the following. 
In contrast, popular ED choices such as Sobol' or Halton low-discrepancy sequences or Latin hypercube sampling (LHS) may exhibit suboptimal projection properties \cite{Sobol:67,Halton:60,McKayConovBeck:three:1979}. 

In addition to data scarcity, a large number of input parameters can further hinder surrogate-based optimization due to the curse of dimensionality. 
The complexity of practical turbomachinery design motivates the assumption of anisotropic parameter impact, meaning that only a subset of the input parameters have a meaningful impact on the response. 
We call the corresponding parameter subspace \emph{influential}, as opposed to the \emph{non-influential} parameters that are omitted therein.
Variance-based GSA provides a practical tool for quantifying the importance of the input parameters through sensitivity metrics known as Sobol' indices \cite{saltelli2008global,Sobol}, and thus identifying an influential subspace of reduced dimensions. 
Polynomial chaos expansion (PCE) surrogate models allow to estimate Sobol' indices cost-efficiently \cite{SUDRET}, which is crucial in this setting. 

Prior works have explored the combination of BO and MaxPro EDs in the broader setting of design and analysis of computer experiments \cite{huang2021bayesian,song2025efficient}, and have linked GSA with MaxPro EDs \cite{cataldo2025global,jivani2023global}. 
The main contribution of this paper is the connection of all three components in an end-to-end optimization workflow, which is able to provide high-efficiency radial turbine designs under strict simulation budget limitations. 
Our numerical experiments clearly demonstrate that combining MaxPro EDs with robust space-filling and projection properties, input dimension reduction based on GSA, and BO performed first in the full parameter space and then in a reduced influential subspace, improves turbine efficiency significantly, while preserving a low computational cost.

The remaining of this work is organized as follows. Section~\ref{sec:problem_setup} introduces the turbine design problem setup, the probabilistic input modeling, and the high-fidelity CFD model with dropwise condensation. 
Section~\ref{sec:methodology} presents the proposed methodology for the end-to-end workflow, including the MaxPro initial ED, surrogate modeling with GPs and PCEs, BO, and PCE-based GSA. 
Numerical results are discussed in Section~\ref{sec:num-results}. 
Last, Section~\ref{sec:conclusion_and_outlook} summarizes the findings of this paper and outlines possible future research directions.

\section{Problem Setup} \label{sec:problem_setup}
Consider a computational model $\mathcal{M}$ in the deterministic sense, mapping from $\mathbb{R}^N$ to $\mathbb{R}$ with
\begin{equation}  
    \mathcal{M}\left(\mathbf{x}\right) = y. \label{eq:computational_model}
\end{equation}
Here, $\mathcal{M}$ coincides with the considered turbine model, the scalar response $y$ is the turbine thermodynamic efficiency $\eta$, and $\mathbf{x} = \left(x_1, \dots, x_{N} \right)$ is a vector of design parameters.  
More details regarding the design parameters and the turbine model are available in Sections~\ref{sec:prelim-turbine-design} and \ref{sec:CFD Model}.

To analyze the computational model under more realistic conditions, input uncertainties must be taken into account. 
A common approach is to model each individual input as a random variable $X_n$ following a suitable marginal probability distribution, with corresponding probability density function (PDF) $f_{X_n}$ and support $\Gamma_{X_n} \subset \mathbb{R}$, such that $f_{X_n}: \Gamma_{X_n} \rightarrow \mathbb{R}_{\geq 0}$.
Accordingly, the joint PDF and support are respectively denoted with $f_\mathbf{X}$ and $\Gamma_{\mathbf{X}} \in \mathbb{R}^N$, where $f_{\mathbf{X}}: \Gamma_{\mathbf{X}} \rightarrow \mathbb{R}_{\geq 0}$.
The corresponding joint cumulative distribution function (CDF) is denoted as $F_{\mathbf{X}}(\mathbf{x}) = \mathbb{P} \left(\mathbf{X} \leq \mathbf{x} \right)$, where $\mathbb{P}\left(\cdot\right)$ denotes the probability of an event.
Then, the deterministic formulation \eqref{eq:computational_model} evolves into a probabilistic one, such that $Y = \mathcal{M}\left( \mathbf{X} \right)$. 
Now, the model's input is the random vector $\mathbf{X}=\left( X_1, \dots, X_N\right)$ defined on a suitable probability space $\left(\Omega, \Sigma, \mathbb{P} \right)$, with sample space $\Omega$, event space $\Sigma$, and probability measure $\mathbb{P}$.  
Accordingly, the response is now a random variable dependent on the input random vector.

Note that the problem setup above and the methodology discussed in Section~\ref{sec:methodology} assume independent input random variables $X_n$, $n=1,\dots,N$. However, dependent inputs can be accommodated with suitable modifications \cite{JAKEMANDependentPCE,chastaing2012generalized}.

\subsection{Preliminary Turbine Design}
\label{sec:prelim-turbine-design}

\begin{figure}[t]
  \centering
  \begin{minipage}[t]{0.56\textwidth}
    \vspace{0pt} 
    \captionof{table}{Turbine operating point parameters.}
    \label{tab:turbine_operating_point}
    \centering
        \begin{tabular}{c c c c}
        \hline
        Parameter & Symbol & Value & Units\\
        \hline
        Total pressure ratio & $\pi_{\mathrm{tt}}$ & $4.42$  & --\\
        Inlet total pressure & $p_{\mathrm{t, in}}$ & $2.08$ & bar    \\
        Inlet total temperature & $T_{\mathrm{t, in}}$ & $ 358.15$ & K \\
        Mass flow rate & $\dot{m}$ & $0.411$ & kg/s\\
        Inlet relative humidity & $\varphi_{\mathrm{t, in}}$ & $0.6$ & -- \\
        \hline
        \end{tabular}
  \end{minipage}\hfill
  \begin{minipage}[t]{0.44\textwidth}
    \vspace{0pt} 
    \centering
    \includegraphics[width=\linewidth]{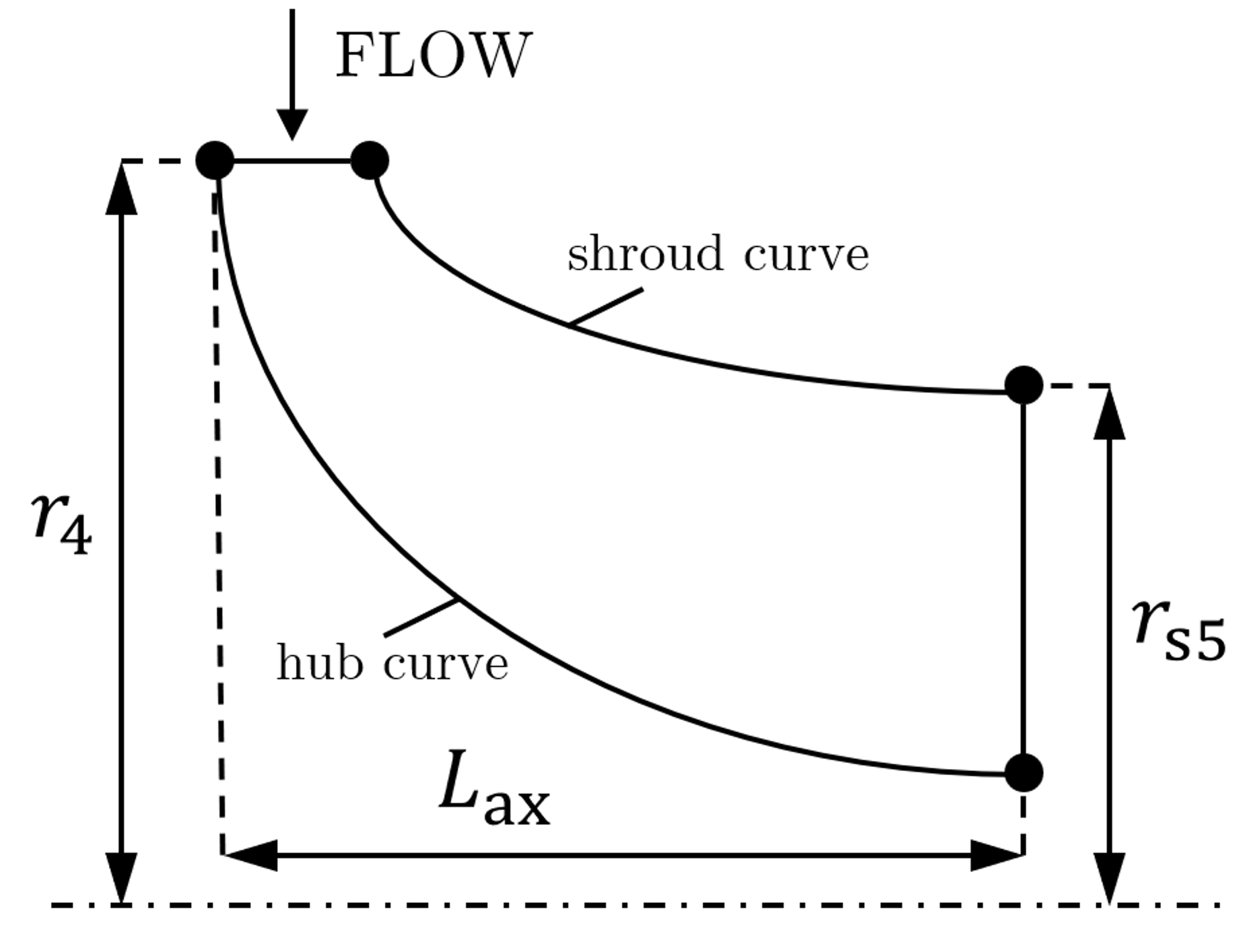}
    \caption{Side view on the turbine rotor flow channel and its parametrization.}
    \label{fig:turbine_annulus}
  \end{minipage}
\end{figure}

The computational model $\mathcal{M}$ from Eq. \eqref{eq:computational_model} represents a single-stage radial turbine in the present work. 
The investigated geometries consist of a stator ring and a turbine rotor. 
The considered operating point is determined by the design of a PEMFC air supply system for aviation applications \cite{lueck_2024}. 
All thermodynamic parameters of the operating point  are summarized in Table \ref{tab:turbine_operating_point}. 
Since only a single operating point is considered in the current study, the stator geometry is determined from a constant parametrization, such that the correct inflow angle for the turbine rotor is induced.
Hence, only the turbine rotor parameters, as introduced below, vary in the computational model.

\renewcommand{\arraystretch}{1.15}
\begin{table}[t]
\caption{Probabilistic input parameters (all dimensionless). The unknown probability distributions for IVR, SS, and FC are initially chosen to be uniform and later refined via parametric probability distribution fitting.}
\label{tab:pre_design_inputs}
\centering
\begin{tabular}{c c c c c}
\hline
$X_n$ & Symbol & Parameter &  Distribution \\
\hline
$X_1$  & IVR & Isentropic Velocity Ratio & Unknown/Data-driven fit  \\
$X_2$  & SS  & Specific Speed            & Unknown/Data-driven fit  \\
$X_3$  & FC  & Flow Coefficient          & Unknown/Data-driven fit \\
$X_4$  & SRR & Shroud Radius Ratio       & Uniform, [0.6, 0.85] \\
$X_5$  & ALR & Axial Length Ratio        & Uniform, [0.3, 0.5] \\
$X_6$  & BN  & Blade Number              & Discrete uniform, \{12--16\} \\
$X_7$  & CAD & {Camber Angle Distribution Shape} & Uniform, [-120, 120] \\
$X_8$  & MT  & Maximum Thickness         & Uniform, [3, 5] \\
$X_9$  & HCS & Hub Curve Shape           & Uniform, [0.6, 1.0] \\
$X_{10}$ & SCS & Shroud Curve Shape      & Uniform, [0.1, 0.3] \\
\hline
\end{tabular}
\end{table}

The preliminary design approach follows a structure similar to the one presented by Ventura et al. \cite{ventura_2012}. Figure \ref{fig:turbine_annulus} shows the parametrization of the turbine rotor flow channel. 
The geometrical dimensions $r_4$, $L_{\mathrm{ax}}$, and $r_{\mathrm{s}5}$ correspond to the rotor radius, axial length, and outlet shroud radius, respectively. 
Inflow and outflow width follow from thermodynamic considerations in order to reach the targeted mass flow. 
The inflow and outflow sections are connected with the hub and shroud curves, which define the endwalls of the axisymmetric flow channel. 

Table \ref{tab:pre_design_inputs} summarizes the $N=10$ parameters that are used to uniquely define the geometry of a specific turbine design. 
The isentropic velocity ratio (IVR) and specific speed (SS) parameters ($X_1$ and $X_2$, respectively) define the main dimension of the turbine rotor, $r_4$, given as
\begin{equation}
r_4 (X_1, X_2) = \left( \sqrt[4]{\frac{4 \dot{m}^2}{\rho_5^2 \Delta h_{\mathrm{t,is}}}} \ \right) \frac{1}{X_1  X_2} ,
\end{equation}
with the outflow density $\rho_5$ and the turbine stage isentropic enthalpy difference $\Delta h_{\mathrm{t,is}}$. 
The rotor outflow velocity $c_5$ influences the outflow area of the rotor and is given as
 \begin{equation}
 c_5(X_1, X_3) = \left( \sqrt[2]{2 \Delta h_{\mathrm{t,is}}} \ \right) \frac{1}{X_1  X_3},
\end{equation}
where $X_3$ corresponds to the flow coefficient (FC) parameter.
The shroud radius at the rotor outlet, $r_{\mathrm{s}5}$, and the axial length of the rotor, $L_{\mathrm{ax}}$, are given as
\begin{align}
 r_{\mathrm{s}5}(X_1, X_2, X_4) &= r_4(X_1, X_2) X_4, \\
 L_{\mathrm{ax}}(X_1, X_2, X_5) &= 2 r_4(X_1, X_2) X_5,
\end{align}
where $X_4$ and $X_5$ correspond to the shroud radius ratio (SRR) and the axial length ratio (ALR) parameters, respectively.
The total number of rotor blades around the circumference is defined by the blade number (BN) parameter ($X_6$). 
The turbine blades are constructed with a quartic polynomial camber angle distribution and a perpendicular thickness distribution. 
The blade camber angles at the inlet and outlet are set by the preliminary design algorithm. 
The distribution of curvature between the end points is parametrized with the camber angle distribution shape (CAD) parameter ($X_7$).
The same non-dimensional camber angle distribution is used at the blade hub and tip. 

The maximum thickness (MT) parameter ($X_8$) sets the maximum value of the quartic thickness distribution at the hub. 
All other parameters of the quartic thickness curve also scale with $X_8$, keeping its non-dimensional shape constant.
A constant thickness is prescribed at the blade tip, which is set to $0.25 X_8$. 
The camber angle and thickness distributions are defined for the hub and the blade tip and then connected linearly. 
The distribution of curvature along the hub end shroud curves is set with the hub curve shape (HCS) and shroud curve shape (SCS) parameters ($X_9$ and $X_{10}$, respectively).

The analytical preliminary design results in an initial 1D description of the turbine flow kinematics and thermodynamics for the hub and the shroud of the turbine. 
The employed design algorithm involves initial guesses for the turbine efficiency, outflow density, and exit flow angles, which will differ from the result of a more sophisticated CFD solution.
Hence, an iterative approach is used to converge the design towards the desired specifications.
Starting from the initial guess of the turbine geometry, a CFD calculation is performed which updates the turbine efficiency, outflow density and exit flow angles. 
These are then fed back as inputs to the preliminary design algorithm. 
The result is an updated geometry, also satisfying the operating point specifications in  Table \ref{tab:turbine_operating_point} more closely.
This procedure is repeated until the design reaches a 1\% error range with respect to the specified mass flow and total pressure ratio.
Using the CFD model with phase change for these design iterations would be very time consuming, especially considering the total number of investigated designs. 
For this reason a simplified CFD model without phase change is used to converge the turbine designs.

Details on the different CFD models used are given in Section~\ref{sec:CFD Model}. 
Once the turbine design is converged, the CFD model with phase change is applied to the resulting geometry. 
The inclusion of phase change shifts the operating point of the turbine, as well as the first three parameters in Table \ref{tab:pre_design_inputs}, by up to 2\%. 
This deviation is accepted due to the substantial time saved by using a simplified CFD model for the preliminary design iterations. 
The complete workflow introduced in Section \ref{sec:methodology} always uses the final output values of the CFD calculations with phase change. 
For that reason, the distributions of the parameters IVR, SS, and FC are not known a priori and need to be derived via distribution fitting.

Turbine designs are evaluated on the basis of their thermodynamic efficiency $\eta$. 
Denoting the power output of the turbine with $\dot{P}$, the efficiency is estimated using the formula
\begin{equation} 
\label{eq:efficiency}
\eta = \frac{\dot{P}}{\dot{m} \, \Delta h_{\mathrm{t, is}}}, 
\end{equation}
which relates the turbine's power output to an idealized isentropic adiabatic expansion process.

\subsection{CFD Model} 
\label{sec:CFD Model}

Numerical simulations are conducted with the commercial flow solver \texttt{Ansys CFX 2025R2}. 
Discretized conservation equations are solved using an element-based finite volume approach. 
In particular, the Favre-averaged Navier-Stokes equations are solved for all simulations in this study \cite{favre_1965}:
\begin{align}
&\text{Continuity:}\notag \\
&\frac{\partial \overline{\rho}}{\partial t}
+ \frac{\partial}{\partial x_j}\left(\overline{\rho}\,\widetilde{\widetilde{u}}_j\right) = 0
\\[1em]
&\text{Momentum:}\notag \\
&\frac{\partial}{\partial t}\left(\overline{\rho}\,\widetilde{\widetilde{u}}_i\right)
+ \frac{\partial}{\partial x_j}\left(\overline{\rho}\,\widetilde{\widetilde{u}}_i\widetilde{\widetilde{u}}_j\right)
= -\frac{\partial \overline{p}}{\partial x_i}
+ \frac{\partial}{\partial x_j}\left(\overline{\tau}_{ij}
- \overline{\rho\,u_i''u_j''}\right)
\\[1em]
&\text{Total Energy:}\notag \\
&\frac{\partial}{\partial t}\left(\overline{\rho}\,\widetilde{\widetilde{h}}_t\right)
+ \frac{\partial}{\partial x_j}\left(\overline{\rho}\,\widetilde{\widetilde{u}}_j \widetilde{\widetilde{h}}_t\right)
=
\frac{\partial \overline{p}}{\partial t}
+ \frac{\partial}{\partial x_j}\left(
\overline{q}_j
- \widetilde{\widetilde{u}}_i\,\overline{\tau}_{ij}
- \overline{\rho\,u_j'' h_t''}
\right),
\label{eq:total_energy}
\end{align}
with density $\rho$, velocity vector $u_i$, pressure $p$, viscous stress tensor $\tau_{ij}$, heat flux vector $q_i$, total enthalpy $h_t$, time $t$ and spatial dimensions $x_i$. 
The notations $\overline{(\cdot)}$,  $\widetilde{\widetilde{(\cdot)}}$ and $(\cdot)''$ indicate ensemble-averaging, Favre-averaging, and Favre fluctuations, respectively. 
Menter’s shear stress transport turbulence model \cite{menter_2012} is used for closure. 
The working medium is humid air, which is treated as an ideal mixture of air and water vapor. 
All simulations are performed in steady state. 
Phase change is calculated based on the thermodynamic model introduced by Tosun et al. \cite{tosun_2024}. 
The respective mass and energy source terms are added to the conservation equations for each cell. 
All phases share the same velocity and pressure field. 
The total energy equation \eqref{eq:total_energy} is only solved for the gaseous phase. 
For the liquid phase, an algebraic relation for small droplets is used to calculate the temperature \cite{gyarmathy_1962}.
Fluid flow boundary conditions are set as constant total pressure, total temperature, and water vapor mass fraction at the inlet, and as an average static pressure at the outlet. 
All walls are modeled as adiabatic and with a no-slip velocity boundary condition. 

As mentioned in Section \ref{sec:prelim-turbine-design}, a simplified version of the CFD model detailed above is used to converge each turbine geometry towards the design specifications. 
This simplified model omits the calculation of phase change and does not introduce any source terms to the conservation equations. 
Hence, no liquid phase needs to be transported for the simplified model, further reducing the computational effort.

\begin{figure}[t]
    \centering
    \includegraphics[width=0.6\linewidth]{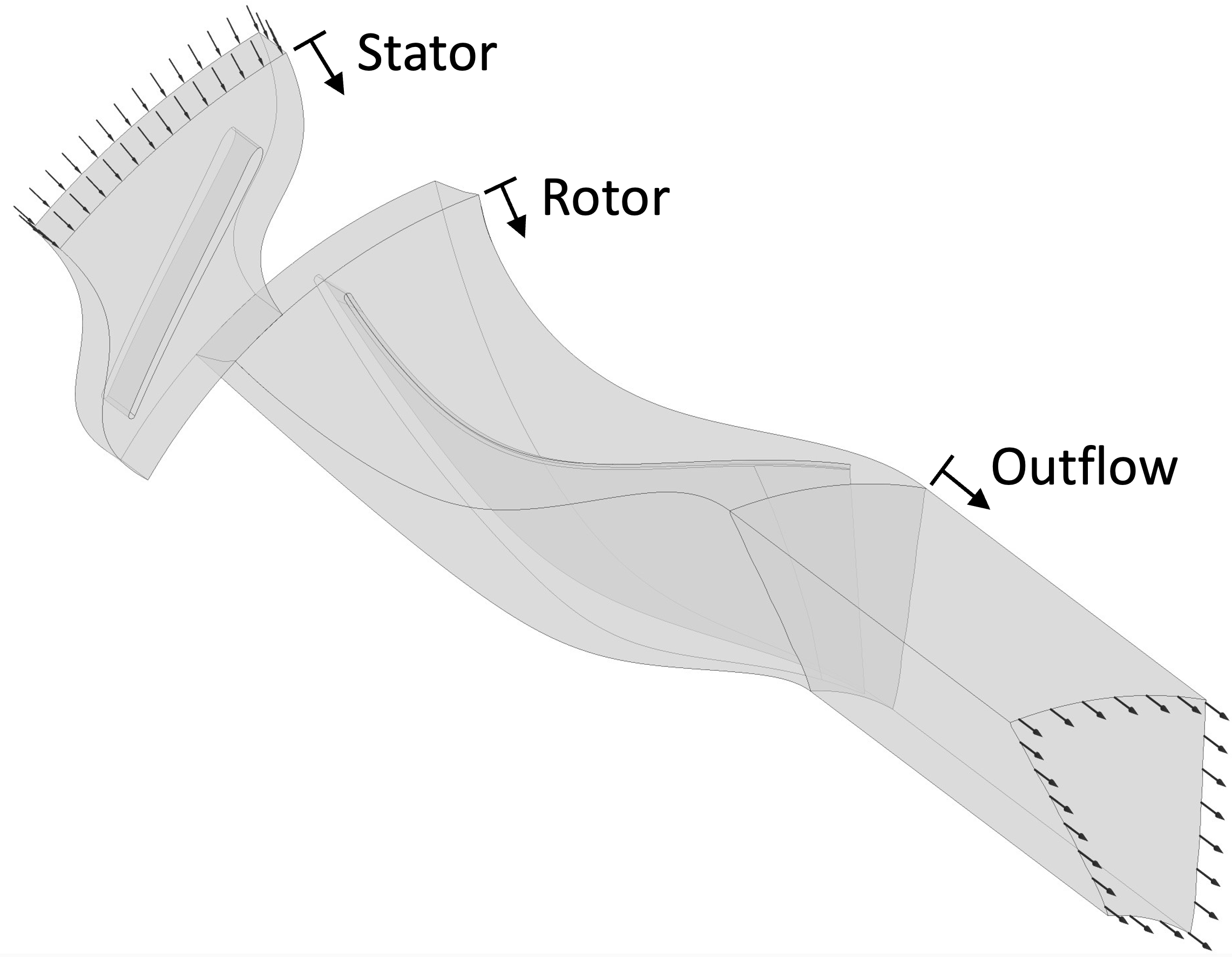}
    \caption{Representative computational domain for the CFD simulation.}
    \label{fig:CFD_domain}
\end{figure}

The computational domain is divided into a stator domain, a rotor domain, and an outflow domain, as shown in Figure \ref{fig:CFD_domain}. 
For the stator and the rotor, only a sector containing a single blade is included in the computational model. 
Periodic boundary conditions are used at each side of the domains, exploiting the rotational symmetry of the geometry. 
The outflow domain has the same circumferential extension as the rotor domain. 
The stator domain and the outlet section downstream of the turbine rotor are solved in a stationary frame of reference. 
The turbine rotor domain is solved in a moving frame of reference, which rotates with the rotor speed. 
The shroud wall of the rotor domain is modeled as a counter-rotating wall, such that it is effectively immobile in the stationary frame of reference.
The interfaces between rotating and stationary domains are modeled with a mixing plane approach \cite{ANSYS_CFX_Solver_Theory_Guide_2025}.

Approximate solutions are obtained on block-structured hexahedral grids with exponential grid refinement towards wall boundaries. 
All computational grids are generated with \texttt{Ansys TurboGrid  2025R2}.
Iterative convergence of the CFD simulations is assessed using the coefficient of variation (CoV) of turbine efficiency 
\begin{equation}
    \mathrm{CoV} = \frac{\sigma_{\eta}(m)}{\mu_{\eta}(m)},
\end{equation}
calculated as the ratio of the efficiency's standard deviation $\sigma_{\eta}(m)$ and the mean efficiency $\mu_{\eta}(m)$ over the last $m=25$ pseudo-timestep iterations.
The convergence criterion is defined as $\mathrm{CoV}<1\cdot10^{-5}$. 
Additionally, the root-mean-square residual norms of the conservation equations are required to be smaller than $1\cdot10^{-4}$. 

\begin{table}[t]
    \caption{Mesh study data and results}
    \label{tab:meshStudyData}
    \centering
    \begin{tabular}{| c | c c c |}
        \hline
         & Mesh 1 & Mesh 2 & Mesh 3 \\
        \hline
        $r_i$ & 1 & 1.98 & 1.98 \\
        $N_i$ & 260,890 & 2,015,418 & 15,608,546 \\
        $\eta_i$ & 0.810187 & 0.830857 & 0.835337 \\
        $e_i$ & 3.18\% & 0.69\% & 0.15\% \\
        $\bar{y}^+$ & 6.42 & 3.52 &  2.01\\
        \hline
    
    \end{tabular}
\end{table}


To assess the discretization error, three homogeneously refined numerical grids are created for a reference turbine geometry. 
Table \ref{tab:meshStudyData} shows the mesh cell count $N_i$, along with the respective refinement factor $r_i$, defined as
\begin{equation}
   r_i = \sqrt[3]{\frac{N_i}{N_{i-1}}}.
\end{equation}
An estimation of the efficiency prediction on an infinitely fine mesh can be obtained using Richardson extrapolation \cite{roache_1997}, such that
\begin{equation}
    \eta_{\text{ex}} = \eta_3 + \frac{\eta_3 - \eta_2}{r_3^p -1},
\end{equation}
where the solutions $\eta_i$ correspond to the respective mesh, as shown in Table \ref{tab:meshStudyData}. 
The approximate order of convergence $p$ is calculated as \cite{celik_2008}
\begin{equation}
    p = \frac{\ln(\frac{\eta_2 - \eta_1}{\eta_3 - \eta_2})}{\ln(r_3)} \approx 2.24.
\end{equation}
From this, the discretization error on meshes 2 and 3 is estimated as
\begin{equation}
    e_i = \frac{1}{\eta_i} \frac{\eta_i - \eta_{i-1}}{r_i^p -1}.
\end{equation}
For mesh 1, the discretization error is estimated as $e_1 = r^p e_2$. 
The numerical uncertainty for a respective solution is usually quantified using the grid convergence index (GCI), where $\mathrm{GCI}_i = 1.25e_i$ \cite{roache_1997}. 
The error approximations can also be found in Table \ref{tab:meshStudyData}. Since the uncertainty band of mesh 2 is already below 1\%, the resolution of this mesh is considered to be sufficient and is used for all numerical results presented in Section~\ref{sec:num-results}. 
Last, the surface-averaged non-dimensional wall distance of wall boundary nodes $\bar{y}^+$ in Table \ref{tab:meshStudyData} indicate that the majority of nodes lie within the viscous sub-layer for mesh 2. 
Hence, the wall boundary layers are mostly resolved with the employed mesh. 

Each iteration in the final simulation setup takes on average $\approx 850$ CPU seconds to calculate. 
The convergence criteria defined above are reached within 250 to 1000 pseudo-time-step iterations, depending on the turbine geometry. 
The corresponding simulation time depends on the allocated resources and architectures. 
These were different between the conducted simulations performed, making a direct estimation of the simulation time to evaluate a turbine design difficult.  
In a first approximation, the simulation time for the high fidelity CFD calculations including phase change may be estimated from the CPU seconds per iteration above, scaled linearly with the available parallelization.
The convergence towards the design specifications with the simplified CFD model took approximately 5 hours on average.


%
\section{Methodology} \label{sec:methodology}
This section describes the proposed end-to-end workflow for generating an efficient turbine design based on scarce data. 
Starting from a suitable initial ED, we construct surrogate models that predict turbine efficiency. 
The surrogate models are then used in two ways: on the one hand, to generate further adaptive samples by evaluating the CFD simulation on designs proposed by BO; on the other hand, to perform variance-based GSA and enable dimensionality reduction of the input space for continued lower-dimensional optimization.

This section is structured as follows. Section~\ref{sec:initial_ed} discusses the initial ED choice, which is based on the MaxPro method \cite{joseph2015maximum}. Section~\ref{sec:surrogate_modeling} introduces the necessary basics for surrogate modeling with GPs and PCEs. 
Section~\ref{sec:bo} describes the BO setup, including how GP inference is used to suggest new samples via an acquisition function \cite{shahriari2015taking} and adaptively enrich the ED. 
Section~\ref{sec:gsa} presents the fundamentals of variance-based GSA \cite{Sobol}, along with the PCE-based estimation of Sobol' sensitivity indices \cite{SUDRET}. 
Section~\ref{sec:workflow} describes the end-to-end workflow.

\subsection{Initial Experimental Design}
\label{sec:initial_ed}
Given the expensive-to-evaluate CFD turbine model and the assumption of anisotropic behavior in the input space, the initial (non-adaptive) ED must ensure not only good global space-filling properties, but also good projection properties, meaning that the ED remains sufficiently space-filling when projected onto a lower-dimensional subspace. 

Given these considerations, full factorial designs are excluded due to their suffering from the curse of dimensionality. 
Monte Carlo sampling can lead to clustering and gaps, which is unfavorable for surrogate modeling, especially given limited data. 
Quasi-Monte Carlo or low-discrepancy sequences like Sobol' or Halton provide good uniformity, however, their projection properties are not necessarily optimal for surrogate-based optimization under anisotropies \cite{Sobol:67,Halton:60}. 
LHS offers an excellent $1$D projection property, but cannot guarantee sufficient space filling in higher-dimensional subspaces \cite{McKayConovBeck:three:1979}. 
Optimized LHS variants based on maximin, minimax, or $\phi_p$ distance criteria improve global space filling \cite{JohMooYlv:MixiMinMiniMax:JSPI:90,MorMit:JStatPLanInf:95}, but may still exhibit poor projection properties or may require non-trivial adjustments, e.g., choosing the $\phi_p$ hyperparameter properly. 
For instance, the classical maximin criterion maximizes the minimum pairwise distance between any two design points, i.e.,
\begin{equation}
\max_{\mathbf{X}\subset(0,1)^N}\; \min_{1\le i<j\le Q}\; \lVert \mathbf{x}_i-\mathbf{x}_j\rVert_2,
\label{eq:maximin}
\end{equation}
where $Q$ denotes the ED size. 
The $\phi_p$ criterion of \cite{MorMit:JStatPLanInf:95} provides a smooth distance-based alternative, typically minimized as
\begin{equation}
\phi_{p}(\mathbf{X})=
\left(
\sum_{1\le i<j\le Q}
\lVert \mathbf{x}_i-\mathbf{x}_j\rVert_2^{-p}
\right)^{1/p},
\qquad p>0.
\label{eq:phip}
\end{equation}
Both criteria primarily promote well-separated points in the full $N$D space through Euclidean distances and do not explicitly target space-filling behavior under projection. In addition, the $\phi_p$ criterion requires selecting the exponent $p$.

For these reasons, we adopt a design criterion that directly targets projection robustness.

The method of choice in this work is MaxPro, the designs of which explicitly aim to maximize projection properties in all lower-dimensional subspaces, making them robust when the influential parameters are unknown, while also being globally space-filling \cite{joseph2015maximum}. 
More precisely, for $Q$ design points $\mathbf{x}_1,\dots,\mathbf{x}_Q \in (0,1)^N$ assembled in a design matrix $\mathbf{X}\in\mathbb{R}^{Q\times N}$, we minimize the MaxPro criterion
\begin{equation}
\Psi_{\mathrm{MaxPro}}(\mathbf{X})=
\left[
\frac{1}{\binom{Q}{2}}
\sum_{1\le i<j\le Q}
\left(
\prod_{n=1}^{N} (x_{in}-x_{jn})^{2}
\right)^{-1}
\right]^{\frac{1}{N}},
\label{eq:maxpro}
\end{equation}
where the product penalizes pairs that are close in any coordinate, hence discouraging clustering in low-dimensional projections.

In contrast to the distance-based criteria in Eqs.~\eqref{eq:maximin}--\eqref{eq:phip}, which enforce separation via full-space Euclidean distances, the MaxPro objective in Eq.~\eqref{eq:maxpro} explicitly couples all coordinate-wise separations through the product term. 
As a result, even if two points are well separated in the full $N$D space, they are strongly penalized if they are close along any single coordinate, which directly promotes robust space filling in projected subspaces.
However, this advantage is achieved at the expense of increased computational cost. 
Generating a MaxPro ED requires solving a ($Q\times N$)-dimensional optimization problem. 
In our implementation, we initialize from a Latin hypercube design and minimize \eqref{eq:maxpro} using a global optimization strategy based on simulated annealing, which is computationally tractable for the present setting ($N=10$, $Q_0=200$).
Therefore, we use MaxPro for the initial ED, providing the foundation for adaptive sampling via BO.

\subsection{Surrogate Modeling}
\label{sec:surrogate_modeling}
To enable data-efficient optimization, the high-fidelity CFD model $\mathcal{M}$ from Eq.~\eqref{eq:computational_model} is replaced with a comparatively cost-efficient but less accurate surrogate model $\widetilde{\mathcal{M}}\left(\mathbf{x}\right) \approx \mathcal{M}\left(\mathbf{x}\right)$. 
In the following, the two complementary surrogate models employed in this work are introduced. 
These are selected according to the requirements of BO and variance-based GSA, namely:
\begin{itemize}
    \item[$\bullet$] For BO, the surrogate model is required to provide not only point predictions but also a measure of predictive uncertainty across the entire input space. GP models are an obvious choice in this context, as they provide a full probabilistic surrogate: they yield a predictive mean and variance at any input point, while also capturing correlations across the input space through a covariance function.
    \item[$\bullet$] For GSA, the surrogate model is required to provide reliable sensitivity information, either by inexpensive repeated evaluations, or through the model representation itself. 
    PCEs are an attractive option, as they encode Sobol' sensitivity information directly into their coefficients.
\end{itemize}

\subsubsection{Gaussian process}
A GP is completely specified by a mean function $\mu(\mathbf{x})$ and a covariance function (or kernel) $k(\mathbf{x}, \mathbf{x'})$. With that, a GP models distributions over functions such that any finite set of function values is jointly Gaussian distributed at every point in the parameter design space $\mathcal{X}$.
Conditioning the prior distribution on the available training data $\{(\mathbf{x}_q , y_q) \}_{q=1}^{Q}$ leads to a posterior predictive distribution at any test point $\mathbf{x}_\star \in \mathcal{X}$. The predicted response on the test point is characterized by the predictive mean
\begin{equation}
    \widetilde{\mathcal{M}}_\mathrm{GP}\left(\mathbf{x}_\star \right) := \mu(\mathbf{x}_\star) = \mathbf{k}_\star^\top(\mathbf{K} + \sigma^2_\epsilon\mathbf{I}_Q)^{-1}\mathbf{y}, \label{eq:gp_sm_mean}
\end{equation}
which serves as the GP point prediction,
and the predictive variance
\begin{equation}
    \sigma^2(\mathbf{x}_\star) = k(\mathbf{x}_\star, \mathbf{x}_\star) - \mathbf{k}_\star^\top(\mathbf{K} + \sigma^2_\epsilon \mathbf{I}_Q)^{-1}\mathbf{k}_\star, \label{eq:gp_sm_variance}
\end{equation}
where $\mathbf{K}\in\mathbb{R}^{Q\times Q}$ is the kernel matrix with entries $\mathbf{K}_{ij} = k(\mathbf{x}_i, \mathbf{x}_j)$, $i,j=1,\dots,Q$, $\mathbf{y}=[ y_1, \dots, y_Q ]^\top$ is the vector of training outputs, $\sigma^2_\epsilon$ denotes the noise variance, and $\mathbf{k}_\star=[k(\mathbf{x}_\star,\mathbf{x}_1),\dots,k(\mathbf{x}_\star,\mathbf{x}_Q)]^\top$ \cite{williams1995gaussian,rasmussen2003gaussian}. 
Note that one could assume noise-free observations obtained from the deterministic CFD solver, i.e., $\sigma^2_\epsilon=0$, as elaborated in \cite{sacks1989design}. 
However, for numerical stability, a very small noise variance $\sigma^2_\epsilon = 10^{-10}$ is included in the Gaussian likelihood, yielding an almost interpolating GP while avoiding ill-conditioning.

The kernels applied in this work are stationary with automatic relevance determination (ARD) lengthscales (different characteristic lengthscale per input dimension), namely the radial basis function (RBF) kernel and Matérn kernels with $\nu \in \{ 3/2, 5/2 \}$ \cite{genton2001classes}. Since we work with a Gaussian likelihood in a scarce data setting, closed-form (or exact) GP inference is feasible and thus employed for training the GP models with the \texttt{Python} library \texttt{GPyTorch} \cite{gardner2018gpytorch}.

\subsubsection{Polynomial chaos expansion}
The PCE is rooted in Norbert Wiener's homogeneous chaos \cite{Wiener_PCE} for Gaussian inputs, but has since then evolved into a popular spectral uncertainty quantification method \cite{GHANEM}, including generalized (non-Gaussian) formulations \cite{Askey}. 
PCEs approximate the model output by a finite series of multivariate polynomials $\Psi$ that are orthonormal with respect to the prescribed joint input PDF $f_\mathbf{X}$, such that
\begin{equation}
    \widetilde{\mathcal{M}}_{\mathrm{PCE}}(\mathbf{x}) =  \sum_{{\boldsymbol{\alpha}} \in \Lambda} c_{{\boldsymbol{\alpha}}} \Psi_{{\boldsymbol{\alpha}}} \left( \mathbf{x} \right) \label{eq:pce_sm}.
\end{equation}
Here, $ c_{{\boldsymbol{\alpha}}} \in \mathbb{R} $ are expansion coefficients, each uniquely identified by a multi-index ${\boldsymbol{\alpha}} \in \mathbb{N}^N_0$ that contains the partial degrees of the corresponding multivariate polynomial $\Psi_{{\boldsymbol{\alpha}}}$. 
The multi-indices form the truncated multi-index set $\Lambda \subset \mathbb{N}^N_0$. 
The PCE coefficients are computed non-intrusively by solving a least-squares regression problem with the available training data.


In the numerical experiments, we utilize isotropic, total-degree (TD) bases, where $\left| \boldsymbol{\alpha}\right| \leq \alpha_{\max}$, and anisotropic (sparse) bases constructed using either least angle regression (LAR) \cite{BLATMANLARS} or a sensitivity-adaptive PCE (SAPCE) algorithm \cite{loukrezis2020robust,loukrezis2025multivariate,DiehlMScThesis}.
Due to the relatively high number of inputs ($N=10$) only linear and quadratic TD bases are considered, i.e., $\alpha_{\max} \in \left\{1,2\right\}$.  
TD-PCEs and SAPCEs are computed with an in-house \texttt{Python} implementation based on \texttt{OpenTURNS} \cite{Baudin2016}. 
LAR-PCEs are constructed using the \texttt{Python} software \texttt{UQpyLab}, derived after the \texttt{MATLAB} software \texttt{UQLab} \cite{UQlab}.

\subsection{Bayesian Optimization} 
\label{sec:bo}
BO is a sequential strategy for optimizing expensive black-box functions using a probabilistic surrogate model, typically a GP, which provides both a mean prediction and a measure of predictive uncertainty across the parameter design space \cite{jones1998efficient,shahriari2015taking}, cf. Eqs. \eqref{eq:gp_sm_mean} and \eqref{eq:gp_sm_variance}. 
This probabilistic information is exploited by an acquisition function that balances exploration (favoring global search in high uncertainty input space areas) and exploitation (favoring local search for high objective values). 
Starting from an initial input-output dataset, BO alternates between (a) fitting the GP on the current training data and (b) selecting the next evaluation points by maximizing the acquisition function, followed by evaluating the expensive black-box model and updating the dataset.
In this work, the upper confidence bound (UCB) acquisition function is employed, given as 
\begin{equation}
    a_{\mathrm{UCB}}(\mathbf{x}) = \mu(\mathbf{x}) + \beta\,\sigma(\mathbf{x}),
    \label{eq:ucb}
\end{equation}
where $\mu(\mathbf{x})$ and $\sigma(\mathbf{x})$ denote the GP's predictive mean and standard deviation, as given in Eqs.~\eqref{eq:gp_sm_mean} and~\eqref{eq:gp_sm_variance}, and $\beta>0$ is an exploration weight parameter adjusted across the different phases of BO.
New CFD evaluations in the BO procedure are obtained for the ED samples 
\begin{equation}
    \mathbf{x}_{\mathrm{next}} \in \arg\max_{\mathbf{x}\in\mathcal{X}} a_{\mathrm{UCB}}(\mathbf{x}).
\end{equation}
The corresponding input-output pairs are added to the input-output dataset.

\subsection{Global Sensitivity Analysis}
\label{sec:gsa}
Parameter importance can be quantified using variance-based GSA, which determines how the uncertainty in the model's response can be apportioned to the uncertain model inputs \cite{saltelli2008global}.
Consider a scalar model response as in Eq.~\eqref{eq:computational_model}, and the probabilistic response $Y=\mathcal{M}(\mathbf{X})$, where input uncertainties are modeled by a random vector $\mathbf{X} = \left( X_1, \dots, X_N\right)$ (cf.~Section~\ref{sec:problem_setup}). Further, assume independent input parameters and a response with a finite second moment, i.e.,
\begin{equation}
    \mathbb{E} \left[Y^2 \right] = \int_{\Gamma_\mathbf{X}} \mathcal{M}^2 \left(\mathbf{x} \right) f_\mathbf{X} \left(\mathbf{x}\right) \, \mathrm{d}\mathbf{x} < \infty. 
\end{equation} 
Then, the variance of the response, denoted as $\mathbb{V}[Y]$, can be decomposed as
\begin{equation}
        \mathbb{V} \left[ Y \right] = \sum_{n=1}^N \mathbb{V}_n + \sum_{n<\bar{n}}^N \mathbb{V}_{n\bar{n}} + \dots + \mathbb{V}_{1,\dots,N}, 
    \label{eq:ANOVA}
\end{equation}
where the term $\mathbb{V}_n$ is the partial variance assigned to the random variable $X_n$ alone; $\mathbb{V}_{n\bar{n}}$  is the partial variance assigned to only combinations of random variables $X_n$ and $X_{\bar{n}}$, with $ n \ne \bar{n}, \, \forall n,\bar{n}=1,\dots,N$; and so on for the remaining partial variances. 
That is, the contribution to the output variance due to a specific uncertain input or combination of inputs is captured by the respective partial variance. 
Then, the first-order and total-order Sobol' sensitivity indices are defined as 
\begin{subequations}
\label{eq:sobol-indices}
\begin{align} 
    S_n^\mathrm{F} &= \frac{\mathbb{V}_n}{\mathbb{V}\left[ Y \right]}, \label{eq:Sobol_Indices_first} \\
    S_n^\mathrm{T} &= \frac{1}{\mathbb{V} \left[ Y \right]} \left( \mathbb{V}_n + \sum_{\substack{ \bar n=1 \\ \bar n\ne n }}^{N} \mathbb{V}_{n\bar{n}} + \dots + \mathbb{V}_{1,\dots, N}\right). 
    \label{eq:Sobol_Indices_total}
\end{align}
\end{subequations}
The first-order Sobol' index quantifies the variance contribution of the input $X_n$ alone, i.e., omitting all interactions with the other inputs. 
The total-order Sobol' index captures the contribution of $X_n$ including all of its interactions with the other inputs.

Sobol' indices can be estimated using Monte Carlo algorithms \cite{saltelli2002making}, however, these can be too computationally demanding, requiring numerous model evaluations. 
Up to moderately high dimensions, PCEs offer a much more efficient alternative \cite{SUDRET}. 
Assuming a PCE in the form of Eq.~\eqref{eq:pce_sm}, Sobol' indices can be estimated by simply post-processing the PCE coefficients, such that
\begin{subequations}
\label{eq:sobol-indices-pce}
\begin{align}
    S_n^{\mathrm{F}} &\approx
    \frac{\sum_{\boldsymbol{\alpha}\in\Lambda_n^{\mathrm{F}}} c_{\boldsymbol{\alpha}}^2}
         {\sum_{\boldsymbol{\alpha}\in\Lambda\setminus\{\mathbf{0}\}} c_{\boldsymbol{\alpha}}^2},
    \qquad
    \Lambda_n^{\mathrm{F}}=\Big\{\boldsymbol{\alpha}\in\Lambda:\alpha_n>0,\ \alpha_{\bar n}=0\ \forall \bar n\neq n\Big\},
    \label{eq:pce_sobol_first}
    \\
    S_n^{\mathrm{T}} &\approx
    \frac{\sum_{\boldsymbol{\alpha}\in\Lambda_n^{\mathrm{T}}} c_{\boldsymbol{\alpha}}^2}
         {\sum_{\boldsymbol{\alpha}\in\Lambda\setminus\{\mathbf{0}\}} c_{\boldsymbol{\alpha}}^2},
    \qquad
    \Lambda_n^{\mathrm{T}}=\Big\{\boldsymbol{\alpha}\in\Lambda:\alpha_n>0\Big\}.
    \label{eq:pce_sobol_total}
\end{align}
\end{subequations}
The common denominator corresponds to the PCE-based approximation of the output variance
\begin{equation}
    \mathbb{V}[Y] \approx \sum_{\boldsymbol{\alpha}\in\Lambda\setminus\{\mathbf{0}\}} c_{\boldsymbol{\alpha}}^2. \label{eq:pce_variance}
\end{equation}
Consequently, the respective numerators of Eqs.~\eqref{eq:sobol-indices-pce} approximate the partial variance components of Eqs.~\eqref{eq:sobol-indices}.

\subsection{Workflow}
\label{sec:workflow}
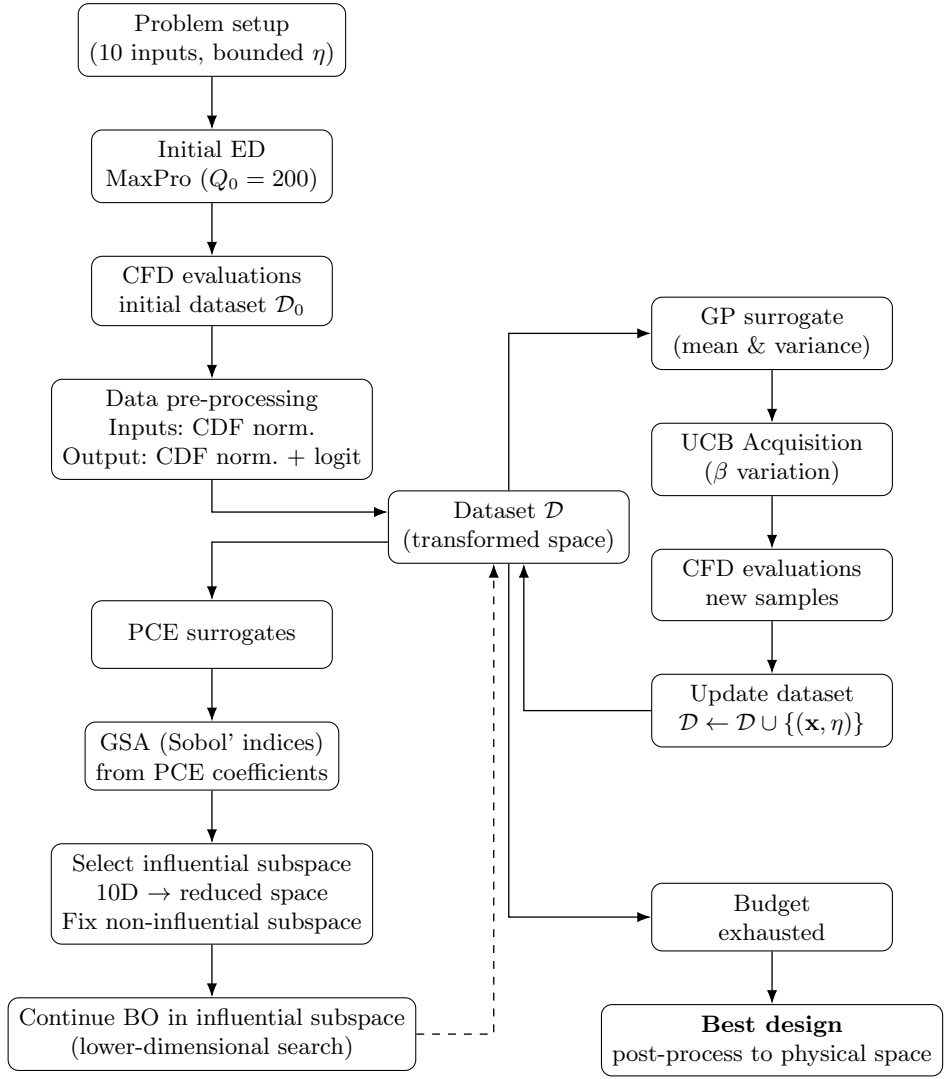
\begin{figure}[!t]
\centering
\begin{tikzpicture}[
  node distance=7mm and 10mm,
  box/.style={draw, rounded corners, align=center, inner sep=4pt, minimum width=3.2cm, minimum height=0.9cm},
  sbox/.style={draw, rounded corners, align=center, inner sep=4pt, minimum width=3.2cm, minimum height=0.9cm},
  arrow/.style={-Latex, line width=0.5pt},
  dashedarrow/.style={-Latex, dashed, line width=0.5pt}
]

\node[box] (setup) {Problem setup\\(10 inputs, bounded $\eta$)};
\node[box, below=of setup] (init) {Initial ED\\MaxPro ($Q_0=200$)};
\node[box, below=of init] (cfd0) {CFD evaluations\\initial dataset $\mathcal{D}_0$};
\node[box, below=of cfd0] (pre) {Data pre-processing\\Inputs: CDF norm.\\Output: CDF norm. + logit};

\node[box, right=2mm of pre, yshift=-13mm] (data) {Dataset $\mathcal{D}$\\(transformed space)};



\node[box, above=16mm of data, xshift=35mm] (gp) {GP surrogate\\(mean \& variance)};
\node[box, below=of gp] (ucb) {UCB Acquisition\\ ($\beta$ variation)};
\node[box, below=of ucb] (cfd) {CFD evaluations\\new samples};

\node[box, below=of cfd] (update) {Update dataset\\$\mathcal{D}\leftarrow\mathcal{D}\cup\{(\mathbf{x},\eta)\}$};

\draw[arrow] (data.north) |- (gp.west);
\draw[arrow] (gp) -- (ucb);
\draw[arrow] (ucb) -- (cfd);
\draw[arrow] (cfd) -- (update);
\draw[arrow] (update.west) -| ([xshift=+2mm]data.south);

\node[box, below=16mm of pre] (pce) {PCE surrogates};
\node[box, below=of pce] (gsa) {GSA (Sobol' indices)\\from PCE coefficients};
\node[box, below=of gsa] (reduce) {Select influential subspace\\10D $\rightarrow$ reduced space\\Fix non-influential subspace};
\node[box, below=of reduce] (bo5d) {Continue BO in influential subspace\\(lower-dimensional search)};

\node[box, below=18mm of update] (budget) {Budget\\exhausted};
\node[box, below=of budget] (best) {\textbf{Best design}\\post-process to physical space};

\draw[arrow] (setup) -- (init);
\draw[arrow] (init) -- (cfd0);
\draw[arrow] (cfd0) -- (pre);
\draw[arrow] (pre.south) |- ([yshift=+2mm]data.west);


\draw[arrow] ([yshift=-2mm]data.west) -| (pce.north);
\draw[arrow] (pce) -- (gsa);
\draw[arrow] (gsa) -- (reduce);
\draw[arrow] (reduce) -- (bo5d);
\draw[dashedarrow] (bo5d.east) -| ([xshift=-2mm]data.south);

\draw[arrow] (data.south) |- (budget.west);
\draw[arrow] (budget) -- (best);
\end{tikzpicture}
\caption{End-to-end workflow. Combines an initial $10$D maximum-projection design, Bayesian optimization (GP modeling and UCB acquisition), and GSA based on PCE for identifying an influential subspace.}
\label{fig:workflow}
\end{figure}
In this section the individual building blocks introduced in Sections~\ref{sec:initial_ed}--\ref{sec:gsa} are merged into an end-to-end workflow. 
Figure~\ref{fig:workflow} provides an overview of the workflow, highlighting practical key mechanisms and serves as visual support for the following step-by-step explanation of the full procedure.
\begin{enumerate}
    \item \textbf{Problem setup and parameter modeling.}
    A 10D parameter design space with a bounded scalar efficiency response is considered. Input marginal distributions are defined as in Table~\ref{tab:pre_design_inputs}, where the a priori unknown marginal distributions need to be fitted based on the data in the current process state.  
    The marginal distributions of the input parameters IVR, SS, and FC are initialized as uniform over physically admissible ranges to generate the initial MaxPro design ($Q_0=200$). After the corresponding CFD evaluations are available, the \emph{realized} parameter values are extracted from the converged CFD results of each evaluated design.  
    These quantities can deviate from their initial targets due to the preliminary-design iteration and the final phase-change CFD evaluation, as also noted in Section~\ref{sec:prelim-turbine-design}. 
    The marginal distributions of IVR, SS, FC, and the output $\eta$ are then \emph{refined} via parametric distribution fitting on the available evaluated designs in the current workflow stage.
    Note that the fitted distribution of $\eta$ is used for the output transformation described in Step~3.
    The fitted distribution is selected using the Bayesian information criterion (BIC) among parametric Normal, Uniform, Beta and Gamma candidate families. For the remaining seven input dimensions, the prescribed uniform distributions in Table~\ref{tab:pre_design_inputs} are used. For the discrete BN parameter, the continuous uniform CDF is used for normalization to maintain a consistent pre-processing across all inputs, while BN is evaluated as an integer in the CFD simulation.

    \item \textbf{Initial maximum-projection experimental design.} An initial MaxPro design with $Q_0 = 200$ samples is generated in the $[0,1]^{10}$-hypercube, transformed to the physical space via inverse CDF transform, and evaluated with the high-fidelity CFD model, yielding the initial input-output dataset $\mathcal{D}_0$.

    \item \textbf{Pre-processing of input and output data.}
    \begin{itemize}
        \item[\textbullet] \textbf{Input data}: The input parameters are normalized with their CDF to a standard uniform space, such that 
        \begin{equation}
           u_n = F_{X_n}(x_n)\in(0,1), 
        \end{equation}
         to eliminate undesired scale-related effects. The resulting dimensionless input space improves surrogate model training and prevents the dominance of high-magnitude variables in sensitivity analysis. 
        \item[\textbullet] \textbf{Output data}: For realistic and practically relevant turbine designs, the efficiency is strictly bounded in $(0,1)$ and exhibits a non-uniform distribution centered around high-efficiency values. Employing statistical models like GPs, which commonly yield normally distributed (unbounded) outputs, the direct modeling of bounded outputs can lead to issues such as predictions outside the valid range. To counteract this undesired behavior,  a two-stage transformation is applied to the output. First, we CDF-normalize the efficiency to obtain a distribution closer to uniform, followed by a logit transformation into an unbounded space. The complete transformation chain is given as 
        \begin{equation}
            u = F_{\eta}(\eta)\in(0,1),  
            \qquad
            z = \mathrm{logit}(u) = \log \left( \frac{u}{1-u} \right) \in \mathbb{R}.
            \label{eq:cdf_logit}
        \end{equation}
        The inverse mapping is accomplished using first the sigmoid function $u=\sigma(z) = \left(1+\exp(-z)\right)^{-1}$, followed by the inverse CDF transform $\eta = F_{\eta}^{-1}(u)$.
    \end{itemize}

    \item \textbf{BO with GP and UCB acquisition function.} A GP model is trained on the current data $\mathcal{D}$ and used within a UCB acquisition function with a varying exploration weight $\beta$. The ED candidate suggested by the acquisition function is then evaluated with the CFD simulation and the resulting input-output pair is appended to dataset $\mathcal{D}$. This procedure is repeated until a sampling or time budget is exhausted. 

    \item \textbf{Variance-based GSA based on PCEs to identify an influential subspace.} In parallel to the BO loop, PCE models are trained to obtain Sobol' indices for the input parameters. Once the sensitivity information is sufficiently stable, the inputs are categorized as either influential (allowed to vary) or non-influential (with fixed values).

    \item \textbf{BO in the reduced (influential) subspace.} The BO loop is continued in the influential subspace, as to allocate the remaining CFD evaluations more efficiently. 
\end{enumerate}
The practical effect of the proposed workflow on efficiency gains, surrogate performance, and sensitivity-driven input dimensionality reduction, is demonstrated in the numerical results presented in the following section.

%
\section{Numerical Results}
\label{sec:num-results}
In this section, the numerical results of the applied workflow are presented, covering surrogate model performance, sequential input-output dataset enrichment during BO, and Sobol' sensitivity analysis results that enable input dimensionality reduction and continued optimization in the identified influential input subspace. 
For clarity, we distinguish between BO \emph{phases}, referring to the exploration-exploitation trade-off induced by the choice of the exploration weight $\beta$ in the UCB acquisition function, and workflow \emph{stages} that describe the major data-generation steps.
The weight values used in BO are $\beta=2$ for the exploration phase, $\beta=1$ for the balanced phase, and $\beta=0.5$ for the exploitation phase.
These workflow stages are:
\begin{enumerate}[label=\text{\textbf{Stage~\arabic*:}}, leftmargin=*, itemsep=0pt, topsep=2pt]
  \item BO with validation in $10$D;
  \item BO without validation in $10$D;
  \item GSA-based input dimensionality reduction from $10$D to $5$D;
  \item BO without validation in $5$D.
\end{enumerate}
\begin{table}[t]
\centering
\small
\setlength{\tabcolsep}{6pt}

\caption{Input-output dataset evolution. The choice of the exploration weight $\beta$ in the UCB acquisition function defines the BO phase, i.e., exploration ($\beta=2$), balanced ($\beta=1$), or exploitation ($\beta=0.5$). The BO batch sizes were adapted to the available compute capacity.}
\label{tab:doe_optimization}
\resizebox{\linewidth}{!}{%
\begin{tabular}{l c l c c c c l}
\toprule
\multirow{2}{*}{Phase} &
\multirow{2}{*}{$\beta$} &
\multicolumn{2}{c}{\makecell{Input-Output \\ Dataset Evolution}} &
\multicolumn{1}{c}{\makecell{Training \\ Dataset}} &
\multicolumn{1}{c}{\makecell{Validation \\ Dataset}} &
\multicolumn{1}{c}{\makecell{Adaptive \\ Dataset}} &
\\
\cmidrule(lr){3-4}
\cmidrule(lr){5-7}
& & Type & \makecell{Batches \\ No.\,$\times$\,Size} &
Size & Size & Size & \\
\midrule

Initial dataset
& -- & MaxPro & $1\times200$ & 200 & 32 & 0
& \multirow{5}{*}{\rotatebox{90}{\scriptsize w/ Validation}} \\

Exploration
& 2 & BO & $1\times15$ & 215 & 32 & 15 & \\

Exploration
& 2 & BO & $5\times5$ & 240 & 32 & 40 & \\

Balanced
& 1 & BO & $3\times5$ & 255 & 32 & 55 & \\

Exploitation
& 0.5 & BO & $3\times5$ & \textbf{270} & 32 & 70 & \\

\addlinespace[0.8ex]
\midrule

\makecell[l]{Add validation \\ samples to \\ training dataset}
& -- & Random & $1\times32$ & 302 & 0 & 70
& \multirow{5}{*}{\rotatebox{90}{\scriptsize w/o Validation}} \\

Exploration
& 2 & BO & $1\times5$ & 307 & 0 & 75 & \\

Exploration
& 2 & BO & $3\times4$ & \textbf{319} & 0 & 87 & \\

$5$D Exploration
& 2 & BO & $3\times3$ & 328 & 0 & 96 & \\

$5$D Exploration
& 2 & BO & $1\times2$ & \textbf{330} & 0 & 98 & \\

\bottomrule
\end{tabular}%
}
\end{table}

Table~\ref{tab:doe_optimization} shows the evolution of the input-output dataset across all workflow stages. 
The sizes of the \emph{training}, \emph{validation}, and \emph{adaptive} (BO-generated) datasets are also documented.
The training dataset is employed to train the GP and PCE surrogate models. 
The validation dataset consists of randomly drawn input-output pairs and is used to evaluate surrogate model quality.
The adaptive dataset size monitors the number of input-output pairs generated by means of BO. 
BO is conducted in batches to reduce time-to-solution, selecting multiple acquisition function suggestions per iteration, according to the available compute resources. 
The number and size of batches are reported in Table~\ref{tab:doe_optimization} as well. 
Table~\ref{tab:doe_optimization} also documents the different BO phases. 

For the batchwise BO samples, a single GP with a Gaussian (RBF) kernel and ARD lengthscales is used to construct the UCB acquisition function. The additional GP kernels (Matérn-$3/2$ and Matérn-$5/2$) are evaluated only as comparative baselines for prediction accuracy on the fixed validation set and are not used to generate BO samples. The RBF kernel choice is motivated by the use of the MaxPro initial design, which exhibits favorable theoretical properties under Gaussian correlation models \cite{joseph2015maximum}.


A detailed description of the numerical results at the different workflow stages follows, where we state the efficiency of the best-so-far turbine design at the end of each stage. 
Finally, a summary of the optimization progress over all four stages is provided.

\subsubsection*{Stage 1: BO with validation in 10D.}
In the first stage, an initial input-output dataset $\mathcal{D}_0$ is obtained by evaluating the CFD turbine model on a MaxPro ED with size $Q_0 = 200$.
The maximum efficiency observed in this initial dataset is
$\eta_{\max}^\mathrm{MaxPro} = 85.77\%$.
Subsequently, the input-output dataset is adaptively enriched by means of RBF-GP-based BO, which is conducted in the full 10D parameter space, up to the dataset size $Q_0 = 270$.
At the end of the first stage, the best turbine design achieves an efficiency $\eta_{\max}^\mathrm{S1} = 89.86\%$, improving the maximum efficiency of the initial dataset by $4.09\%$.

During its adaptive enrichment, the input-output dataset is used to train the GP and PCE surrogate models.
To quantify surrogate model performance, the mean absolute percentage error (MAPE) and the maximum absolute percentage error (MaxAPE) metrics are employed, defined as,
\begin{equation}
\mathrm{MAPE} = \frac{100}{Q_{\mathrm{val}}}\sum_{i=1}^{Q_{\mathrm{val}}}\left|\frac{y_i-\widetilde{y}_i}{y_i}\right|,
\qquad
\mathrm{MaxAPE} = 100 \max_{i=1,\dots,Q_{\mathrm{val}}}\left|\frac{y_i-\widetilde{y}_i}{y_i}\right|,
\end{equation}
where $y_i=\mathcal{M}(\mathbf{x}_i)$ and $\widetilde{y}_i=\widetilde{\mathcal{M}}(\mathbf{x}_i)$. 
These errors are computed using a fixed validation dataset $\mathcal{D}_\mathrm{val}=\{(\mathbf{x}_i,y_i)\}_{i=1}^{Q_{\mathrm{val}}}$ with size $Q_\mathrm{val}=32$. 
We use random samples as validation data to reflect, as closely as possible given the limited data availability, the scenario of predicting the efficiency on unseen designs. 
Note that cross-validation error metrics are not applicable here, since structured space-filling designs are used. 
That is, cross-validation splits would induce structure-altering disruptions in the dataset, hence the need for a separate validation set.

\begin{figure}[!t]
    \centering
    \includegraphics[width=0.75\linewidth]{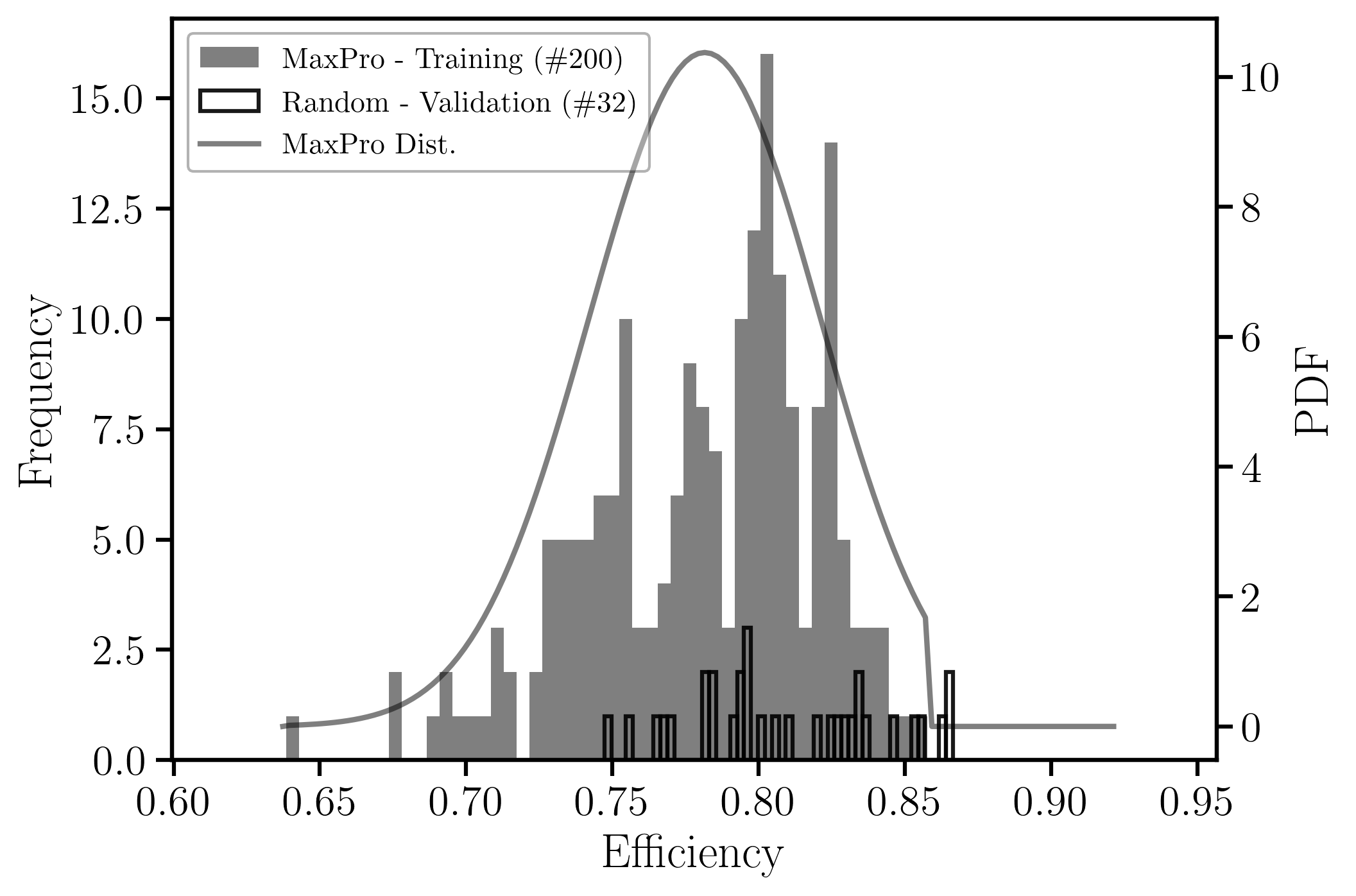}
    \caption{Efficiency distribution of the initial dataset. The gray-marked histogram shows the $Q_0=200$ MaxPro samples, while the black-marked histogram shows the $Q_\mathrm{val}=32$ randomly drawn validation samples. The smooth curve represents a density estimate from the MaxPro dataset.}
    \label{fig:eff_init_histo}
\end{figure}
\begin{figure}[!t]
    \centering
    \includegraphics[width=0.95\linewidth]{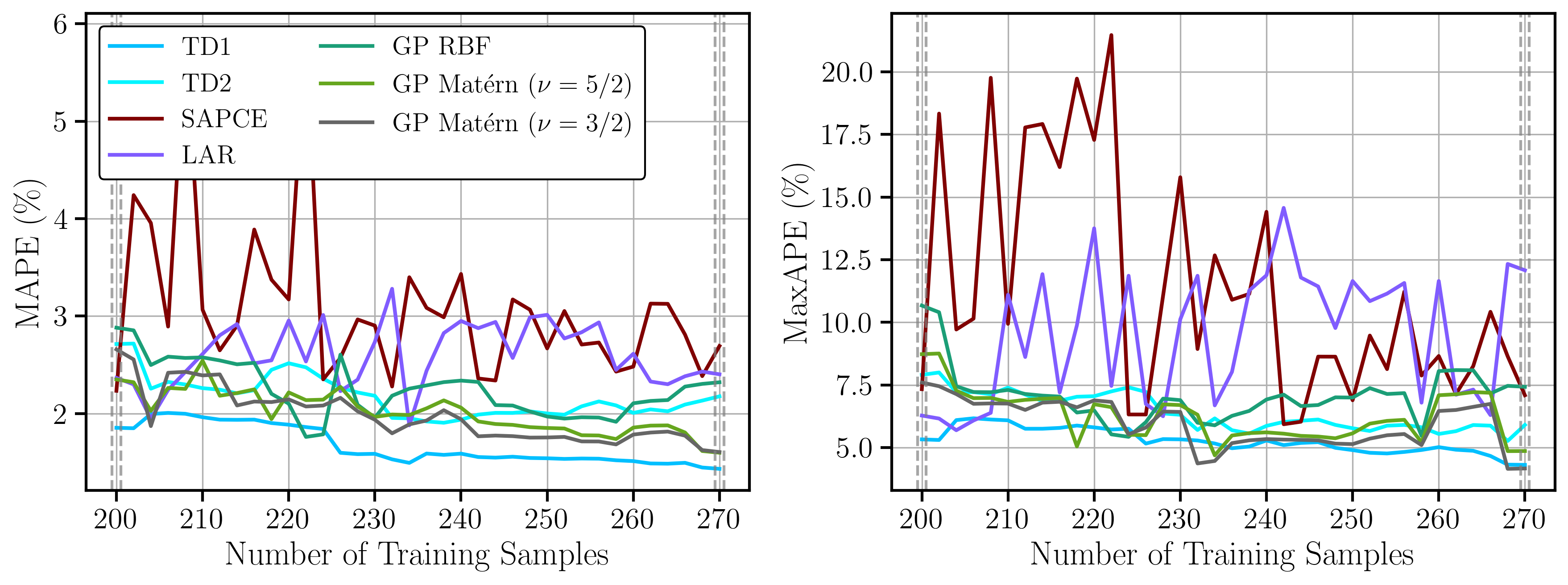}
    \caption{Surrogate model prediction accuracy during stage 1.}
    \label{fig:surrogate_pred_acc}
\end{figure}

The efficiency samples of the initial input-output dataset are shown in Figure~\ref{fig:eff_init_histo}, together with the samples that belong to the validation dataset. 
Surrogate model performance in terms of prediction accuracy during the enrichment of the training dataset is summarized in Figure~\ref{fig:surrogate_pred_acc}.
Overall, linear PCE (TD1) and Mat\'ern-GP models yield the lowest MAPE and MaxAPE, attaining MAPEs below $2\%$ and MaxAPEs below $5\%$ for $Q=270$ training samples, respectively. 
Quadratic PCE (TD2) and RBF-GP models are slightly worse, while the anisotropic PCEs (SAPCE, LAR) yield noticeably larger and more fluctuating error values.

\subsubsection*{Stage 2: BO without validation in $10$D.}

Stage 2 begins once surrogate model performance is sufficiently stable. 
The $32$ random samples used for validation in the previous stage are now added to the training dataset (cf. Table~\ref{tab:doe_optimization}). 
Dataset enrichment by means of BO proceeds once again considering the full $10$D input parameter space.
After two further BO exploration phases leading to a total of $Q=319$ input-output samples, the best turbine design at the end of the second stage reaches an efficiency of $\eta^\mathrm{S2}_\mathrm{max} = 91.53\%$, improving the maximum efficiency by an additional $1.67\%$.
\subsubsection*{Stage 3: GSA-based input dimensionality reduction from $10$D to $5$D.}
\begin{figure}[!t]
    \centering
    \includegraphics[width=0.95\linewidth]{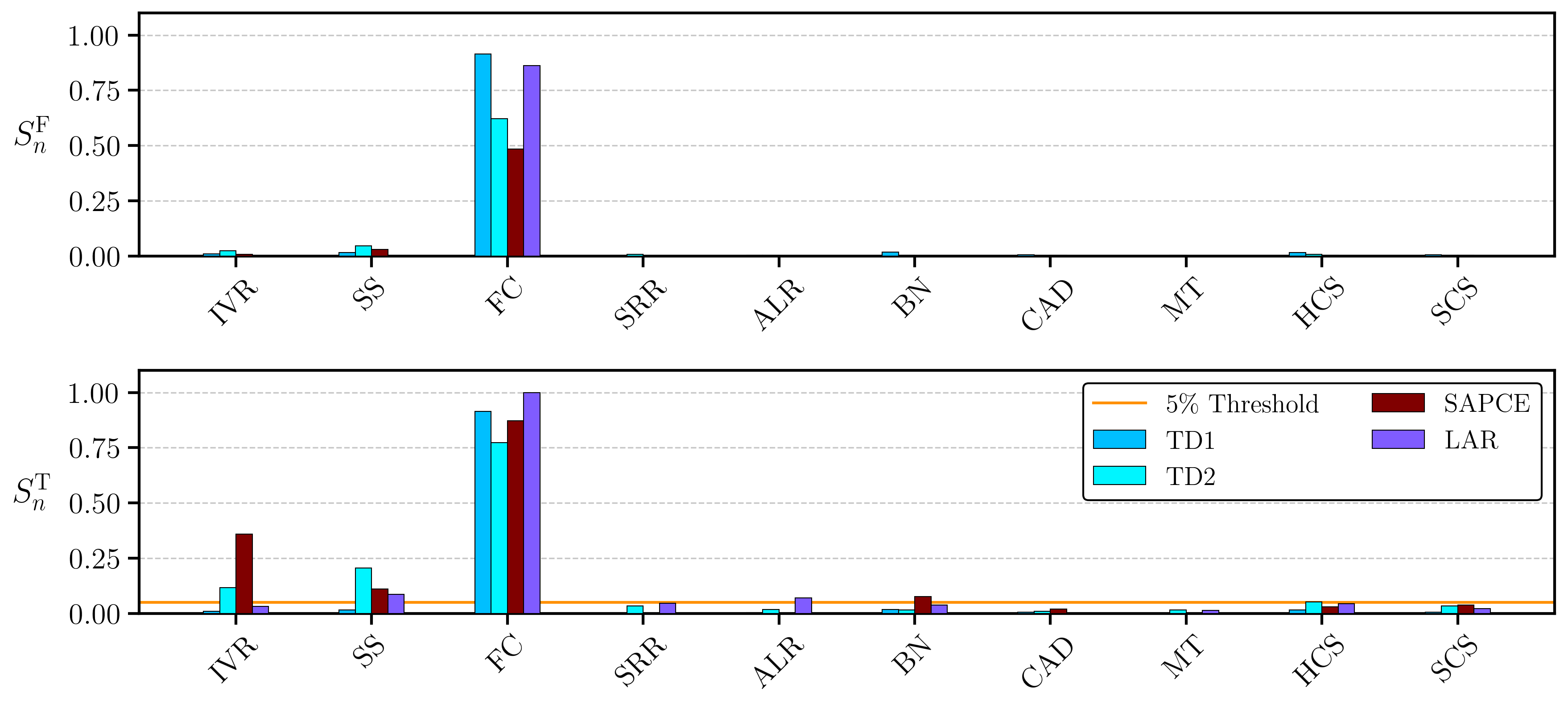}
    \caption{First-order (top) and total-order (bottom) Sobol' indices computed using PCEs trained with 319 samples. The horizontal line indicates the $5\%$ sensitivity threshold used to classify inputs as influential (retained for subsequent optimization) or non-influential (fixed). The final selection is based on the TD2 and SAPCE sensitivity estimates.}
    \label{fig:sens_ana}
\end{figure}
%
%
%
%
In the third stage, variance-based GSA is applied to identify an influential low-dimensional subspace, wherein BO shall be continued. 
Sobol' indices are obtained by post-processing PCE coefficients corresponding to the CDF-normalized and logit-transformed response, cf.~Eq.~\eqref{eq:cdf_logit}.

Figure~\ref{fig:sens_ana} reports first-order and total-order Sobol' indices. 
The differences between first- and total-order indices indicate parameter interaction effects; especially the differences observed for the first two inputs (IVR and SS) indicate that interaction contributions are present and must be accounted for in dimensionality reduction. 
Therefore, the influential subspace selection is based on the total-order indices, as they capture both first-order contributions and all higher-order interactions among the input parameters. 

To distinguish between influential and non-influential parameters, we rely on PCE-based sensitivity estimates obtained from TD2 and SAPCE and apply a $5\%$ total-order sensitivity threshold. 
TD1 and LAR are not used to drive the dimensionality reduction for the different reasons. 
First, TD1 is by construction limited to linear terms and thus cannot represent higher-order interactions, which are relevant here. 
Second, LAR-based sensitivity estimates classify IVR as non-influential, which conflicts with established turbomachinery design knowledge that regards IVR as an important design variable. 
To avoid fixing a physically relevant parameter based on sensitivity estimates that appear inconsistent with domain knowledge in the scarce-data regime, LAR is not used to inform the dimensionality reduction.

Considering the sensitivity threshold of $5\%$ to distinguish influential from non-influential parameters (cf. Figure~\ref{fig:sens_ana}), the parameters IVR, SS, and FC are identified as the dominant variance contributors, while the indices of SRR, ALR, CAD, MT, and SCS remain below the threshold. 
The indices of the remaining two parameters, BN and HCS, are slightly above the threshold for at least one of the two  PCE models. 
To avoid neglecting potentially relevant effects, both parameters are considered influential. 
Consequently, we define the influential $5$D subspace as $\{ \mathrm{IVR},\mathrm{SS}, \mathrm{FC}, \mathrm{BN}, \mathrm{HCS} \}$. 
All non-influential inputs are fixed to the sample mean computed from the ten best-performing designs in the available dataset.

\subsubsection*{Stage 4: BO without validation in $5$D.}
In the fourth and final stage, BO is continued in the influential $5$D subspace ($\{ \mathrm{IVR},\mathrm{SS}, \mathrm{FC}, \mathrm{BN}, \mathrm{HCS} \}$) up to the final budget of $Q=330$ input-output samples (cf. Table~\ref{tab:doe_optimization}). 
At the end of the optimization procedure, the best turbine design yields an efficiency of  $\eta^{\mathrm{S4}}_\mathrm{max} = 91.77\%$.

\paragraph{Remark:} Practically, continuing the BO in the influential subspace only, makes more efficient use of the remaining evaluation budget. 
By fixing the five non-influential inputs, the effective input dimensionality is reduced, which improves sample efficiency and stabilizes the GP surrogate and the acquisition function optimization in the scarce-data regime. Empirically, this benefit is reflected by the increase of the best-so-far efficiency from $91.53\%$ at the end of stage 2, to $91.77\%$ after only $11$ additional BO evaluations in the $5$D subspace at the end of stage 4. 
Continuing the BO in the full space could in principle achieve a similar maximum efficiency, since the influential $5$D subspace is contained in the full space. 
However, since five out of ten parameters contribute only negligibly, exploring the full space would result in elevated computational cost in terms of model evaluations for a comparable improvement.

\subsubsection*{Optimization progress summary.}
\begin{figure}[!t]
    \centering
    \includegraphics[width=0.75\linewidth]{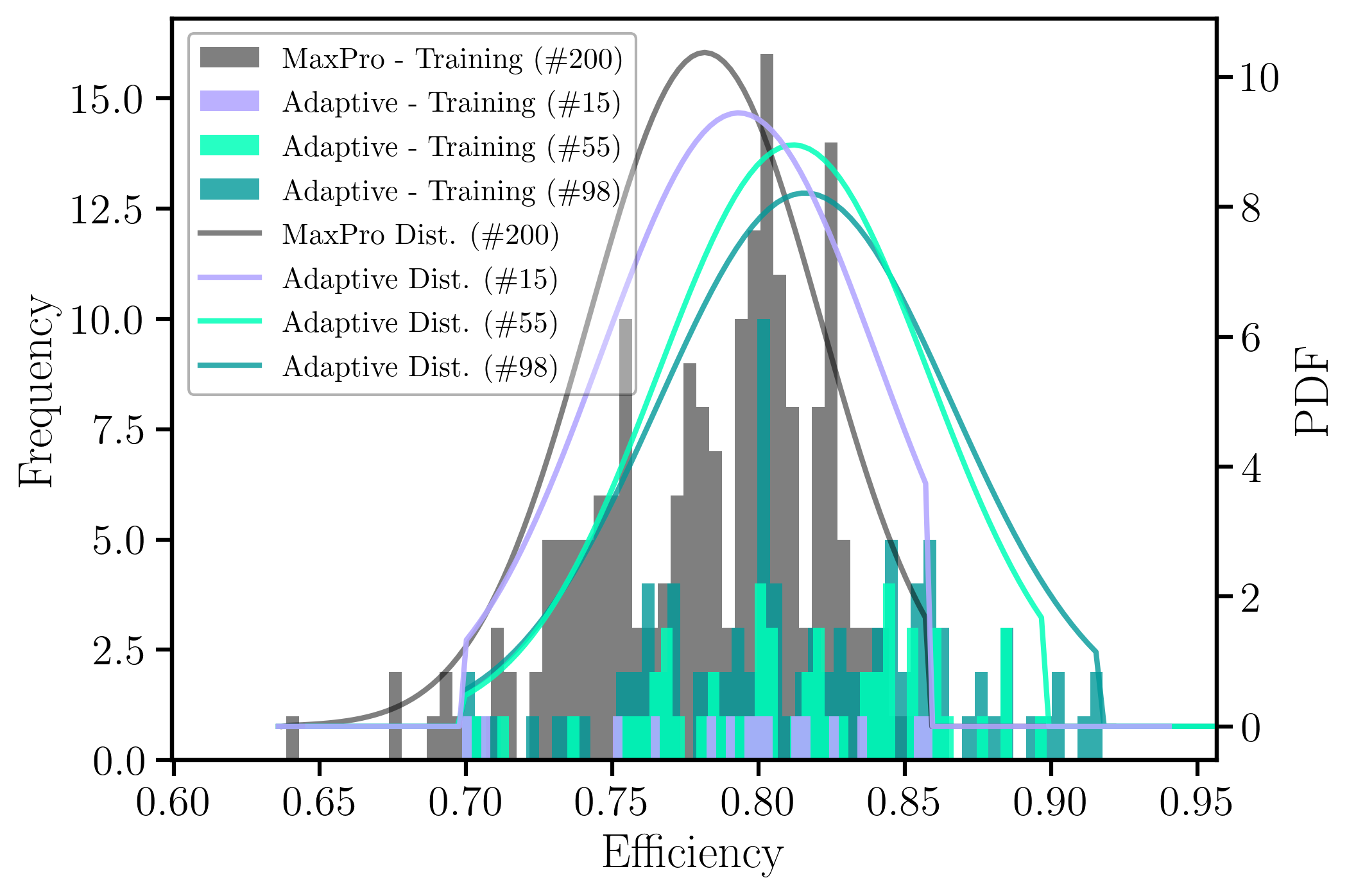}
    \caption{Evolution of the efficiency distribution during adaptive sampling. Histograms show the maximum projection training set ($Q_0=200$) and the added adaptive training samples after selected Bayesian optimization iterations ($Q_\mathrm{adapt}=15,\,55,\,98$). The corresponding smooth curves depict density estimates of the respective adaptive-sample distributions, illustrating the shift of the sampled efficiencies toward higher values (see also Figure~\ref{fig:eff_development}).}
    \label{fig:eff_adapt_histo}
\end{figure}
\begin{figure}[!t]
    \centering
    \includegraphics[width=0.75\linewidth]{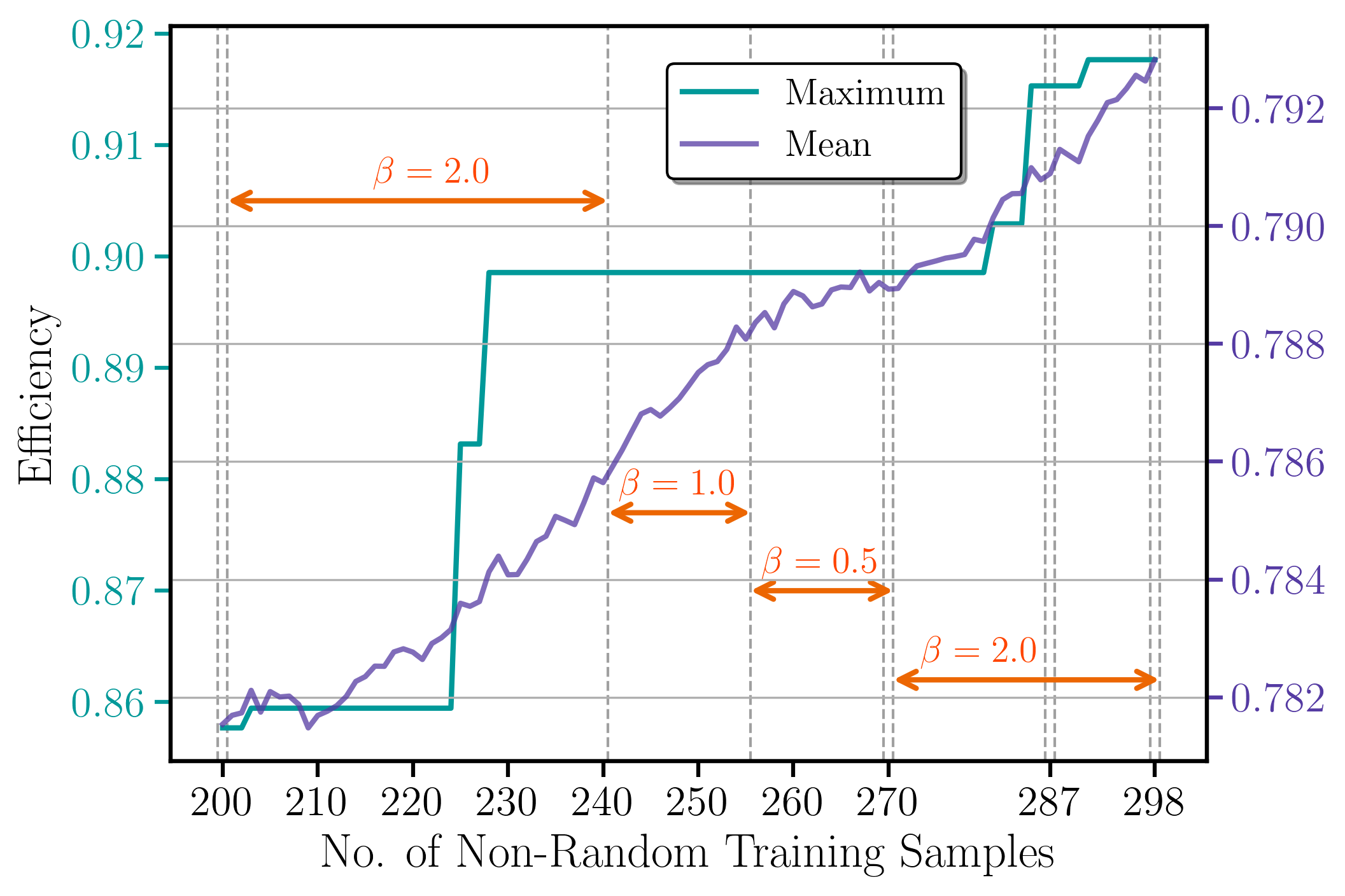}
    \caption{Efficiency evolution during the optimization workflow. Best-so-far (\emph{maximum}) and running mean (\emph{mean}) efficiencies are shown over the number of non-random training samples. Vertical dashed lines indicate the phase boundaries of the BO schedule (cf.~Table~\ref{tab:doe_optimization}). BO is performed in the full $10$D parameter space up to $287$ samples and continued in the reduced $5$D influential subspace up to $298$ samples.}
    \label{fig:eff_development}
\end{figure}
Figures~\ref{fig:eff_adapt_histo} and \ref{fig:eff_development} summarize the overall optimization progress across all workflow stages, considering only the non-random (MaxPro and BO-based) samples, in terms of the sampled efficiency distributions and the evolution of the best-so-far and mean efficiencies.
Figure~\ref{fig:eff_adapt_histo} illustrates how the adaptive samples shift progressively toward higher efficiencies compared to the initial MaxPro dataset, indicating an increasingly exploitative sampling behavior as BO progresses. 
Complementarily, Figure~\ref{fig:eff_development} reports the best-so-far efficiency (\emph{maximum}) and the running mean (\emph{mean}) over the number of non-random samples, with phase boundaries marked according to the BO schedule in Table~\ref{tab:doe_optimization}. 
The figure confirms the consistent improvement of both average and peak performance throughout the workflow. 

Additionally, Figure~\ref{fig:eff_development} shows a pronounced plateau of the maximum efficiency from 228 to 280 non-random $10$D samples, during which no improvement in the best-so-far efficiency is observed. In contrast, after the transition to the reduced $5$D influential subspace at $287$ non-random samples, the final improvement in best-so-far efficiency is achieved with the remaining few evaluations. 
This result is a further indication that higher sample efficiency is achieved when the search is concentrated on influential parameters only.

%
\section{Conclusion and Outlook} \label{sec:conclusion_and_outlook}
This work presented a surrogate-based optimization workflow for data-scarce radial turbine design. 
The design employs computationally expensive $3$D computational fluid dynamics simulations that accounts for dropwise condensation. 
The workflow combines a space-filling maximum-projection experimental design with Bayesian optimization and variance-based global sensitivity analysis to reduce the design parameter space and hence mitigate the computational burden. 
The central objective was to achieve data-efficient optimization under a considerably low budget of high-fidelity simulations. 
The presented numerical results demonstrate that the proposed workflow is able to achieve high-efficiency designs within the considered evaluation budget, while simultaneously providing sensitivity information that enables dimensionality reduction. In particular, the maximum turbine efficiency was increased from $85.77\%$ in the initial, maximum-projection experimental design to $91.77\%$ at the end of the workflow.

A key aspect of the workflow is the complementary use of surrogate models for the different tasks. 
Gaussian process surrogate models are used within Bayesian optimization to provide uncertainty-aware acquisition guidance, while polynomial chaos expansions are used to obtain Sobol' sensitivity indices with low computational demand. 
To support reliable decision-making, surrogate model performance is monitored throughout the early stages in terms of prediction accuracy on a fixed validation set, indicating when the surrogates have reached sufficiently stable behavior. 
Once stable surrogate model performance is established, the validation samples are added to the training dataset to maximize the available information for the final sensitivity-based influential-subspace identification and the remaining optimization iterations.

Several directions for future work can be drawn. Extending the approach to multi-objective optimization would increase practical relevance by involving additional objectives like structural integrity, turbine mass, and inertia. 
The inclusion of multiple operating points could be beneficial as well \cite{lueck_2024}. 
Moreover, since the modular design of the workflow allows to replace individual surrogate components, promising alternative options for probabilistic surrogate models can be explored. 
Potential candidates could be polynomial-chaos-Kriging models that combine polynomial chaos expansion and Gaussian processes \cite{schobi2015polynomial}, or surrogate models augmented with predictive uncertainty estimates based on conformal prediction \cite{fontana2023conformal}. 
In the latter case, conformal prediction methods specifically tailored to polynomial chaos expansions would be of interest \cite{loukrezis2025conformalized,hatstatt2026conformal}.


\subsection*{CRediT authorship contribution statement}
\textbf{Eric Diehl:} Writing – original draft, Writing – review \& editing, Conceptualization, Methodology, Software, Validation, Formal analysis, Visualization. 

\noindent
\textbf{Adem Tosun:} Writing – original draft, Writing – review \& editing, Conceptualization, Methodology, Software, Validation, Investigation, Visualization. 

\noindent
\textbf{Dimitrios Loukrezis:} Writing – original draft, Writing – review \& editing, Validation, Supervision.

\subsection*{Declaration of Generative AI and AI-assisted technologies in the
writing process}

The authors used Gemini 2.5 for text improvements. The authors reviewed and edited all changes and take full responsibility for the final version.

\subsection*{Declaration of competing interest}

The authors declare that they have no known competing ﬁnancial interests or personal relationships that could have appeared to influence the work reported in this paper.


\subsection*{Data availability}
Data can be made available by the authors upon reasonable request. 

\bibliographystyle{unsrt}
\bibliography{bibliography}